\documentclass{article}
\usepackage{blindtext}
\usepackage[a4paper, top=2cm, bottom=3cm, left=1.5cm, right=1.5cm]{geometry}

\usepackage[english]{babel}
\usepackage[x11names]{xcolor} 

\usepackage{amsmath}
\usepackage{makecell}
\usepackage{booktabs}
\usepackage{threeparttable}
\usepackage{multirow}
\usepackage{float}
\usepackage{graphicx}
 \usepackage{cite}
\usepackage[colorlinks=true, allcolors=blue]{hyperref}
\usepackage{algorithm}

\usepackage{bm}
\usepackage{booktabs}
\usepackage{pifont}
\usepackage{tikz}
\providecommand{\keywords}[1]
{
  \small	
  \textbf{\textit{Keywords---}} #1
} 
\begin{document}
\title{Efficient Image Registration for Ultrasound Localization Microscopy by Obtaining Gradients via Integration Across Iterations}

\author{Jipeng Yan, Chang Liu, Hengchang Liu,  Biao Huang, Meng-Xing Tang, Yingxiang Liu, Ying Tan}
\footnotetext[1]{This work was supported by National Nature Science Foundation of China (grant number 62501193, 52225501), Self-Planned Task (No. SKLRS202610C) of State Key Laboratory of Robotics and Systems (HIT), and the National Institute for Health Research i4i (grant number NIHR200972). Jipeng Yan, Chang Liu and Hengchang Liu contributed equally to this work. (\emph{Corresponding authors: Jipeng Yan, Ying Tan, Yingxiang Liu.})} 
\footnotetext[2]{Jipeng Yan was with Ultrasound Lab for Imaging and Sensing, Department of Bioengineering, Imperial College London, London SW7 2AZ, UK, and is with the State Key Laboratory of Robotics and Systems, Harbin Institute of Technology, Harbin 150001, China. (e-mail: jipengyan@hit.edu.cn).}
\footnotetext[3]{Chang Liu and Yingxiang Liu are with State Key Laboratory of Robotics and Systems, Harbin Institute of Technology, Harbin 150001, China. (e-mail: 24S008015@stu.hit.edu.cn, liuyingxiang868@hit.edu.cn)}
\footnotetext[4]{Hengchang Liu and Ying Tan are with School of Electrical, Mechanical and Infrastructure Engineering, The University of Melbourne, Parkville, VIC 3010, Australia.   (e-mail:  hengchang.liu1@unimelb.edu.au, yingt@unimelb.edu.au)}
\footnotetext[5]{Biao Huang and Meng-Xing Tang are with Ultrasound Lab for Imaging and Sensing, Department of Bioengineering, Imperial College London, London SW7 2AZ, UK. (e-mail:  b.huang21, mengxing.tang, @imperial.ac.uk)}

\date{}
\maketitle

\begin{abstract}
Tissue motion correction through image registration is essential for ultrasound localization microscopy (ULM). Parametric image registration is commonly formulated as an optimization problem where motion parameters are iteratively updated to maximize image similarity, and used optimization algorithms typically rely on gradient information, the explicit evaluation of which can become computationally demanding. This work investigates Extremum Seeking Control (ESC) as an alternative to explicit derivative evaluation in image registration. By obtaining descent information via integrating perturbed and demodulated image similarity metric across iterations, ESC avoids differentiation of the image similarity metric with respect to motion parameters in each iteration. The classical ESC, whose optimization behavior approximates that of classical gradient descent (GD), is first compared with GD for affine image registration using simulated ground-truth motions derived from a beating \emph{ex vivo} porcine heart dataset. The results show that ESC achieves registration accuracy and convergence behavior comparable to GD while reducing per-iteration computational cost by approximately 3.5-fold. ESC is subsequently employed in a two-stage motion correction pipeline, where affine registration compensates for global tissue motion and B-spline registration corrects residual local deformation. The proposed method is applied to ULM imaging of a beating \emph{ex vivo} porcine heart and achieves a spatial resolution of 219 $\mu m$, substantially below the half-wavelength diffraction limit of 321 $\mu m$ associated with 2.4 MHz diverging-wave imaging. These results demonstrate that ESC provides an effective alternative to explicit derivative evaluation in ULM image registration, enabling accurate motion correction and high-quality super-resolution imaging.

\end{abstract}

\section{Introduction}

Ultrasound localization microscopy (ULM) enables super-resolution vascular imaging beyond the acoustic diffraction limit by localizing and accumulating signals from individual microbubbles (MBs) \cite{errico2015ultrafast, christensen2014vivo}. By tracking MBs over time, ULM can visualize microvascular structures and quantify blood-flow dynamics at resolutions substantially finer than those achievable with conventional ultrasound imaging, enabling a wide range of preclinical and clinical applications \cite{christensen2020super, song2023super,smith2026quantitative}. However, MB motion reflects not only blood flow but also tissue motion. Without adequate motion compensation, tissue displacement degrades localization accuracy, biases flow estimation, and reduces the achievable ULM resolution. Accurate tissue motion correction is therefore a critical requirement for ULM, particularly in applications involving substantial physiological motions such as heart beating \cite{cormier2021dynamic,demeulenaere2022coronary,yan2024transthoracic}, lung breathing \cite{taghavi2021vivo} and pulsatility in large vessels \cite{demene2021transcranial}.

Parametric image registration has demonstrated compelling performance in estimating complex motions from ultrasound images \cite{dencks2025review_super}.
Among the available motion-correction techniques for ultrasound sequences, the phase correlation method, estimating motion by locating the Dirac peak in the inverse Fourier transform of the cross-correlation between two frames, is limited to estimating translational motion\cite{hingot2017subwavelength};
 Doppler-based motion estimation, inferring tissue displacement from phase shifts in the ultrasound signal \cite{cormier2021dynamic}, can be biased when tissue motion is over half a wavelength; speckle tracking or optical flow methods estimate motion by tracking local speckle patterns \cite{taghavi2021vivo} or features extracted from speckles \cite{wei2024pyramidal}, but ultrasound speckle patterns can  be changed by tissue motions significantly; iterative closest point (ICP) algorithms, estimating deformation with cloud of MB positions \cite{demeulenaere2022coronary}, may be susceptible to the randomness in MB positioning -- particularly when MB concentration is low and the data acquisition time is short.
 Image registration provides an alternative approach by estimating tissue motion through optimization of image similarity between ultrasound images \cite{sotiras2013deformable}. In particular, parametric image registration, where tissue motion is represented by a set of parameters in appropriate models, can accommodate global transformation or local tissue deformation when the discrepancy in motion between ultrasound image patterns and actual tissue arises from speckle and wave-interface effects. Two-stage frameworks \cite{harput2018two}, combining a global transformation model (rigid or affine) with a local deformation model (such as the B-spline based \cite{rueckert2002nonrigid}), have been successfully applied
to challenging applications such as beating-heart imaging \cite{yan2024transthoracic}.

Despite their effectiveness, parametric image registration methods remain computationally expensive because motion parameters in the transformation and deformation models are conventionally estimated through iterative optimization \cite{sotiras2013deformable}. A major source of this computational cost is the repeated acquisition of gradient information required to update motion parameters. In image registration, such gradients are commonly obtained through finite-difference approximations or automatic differentiation, requiring differentiation through image transformation/deformation, interpolation operations, and image-similarity calculations throughout the optimization process. As image resolution and parameter dimension increase, the computational burden associated with gradient evaluation becomes increasingly significant, particularly for deformable registration models containing a large number of parameters \cite{sotiras2013deformable}.  Learning-based techniques can avoid the iterative optimization and result in a significant speed advantage over traditional optimization-based methods. However, supervised learning-based methods are usually trained with the ground truth generated by traditional methods and cannot outperform the traditional methods, and a trained model struggles to perform as expected when presented with input images from a different distribution than the training data \cite{chen2025survey}. As
high precision of motion correction is required and
 various ultrasound devices are used in ULM imaging, improving the computational efficiency of optimization-based image registration remains an important problem for ULM.

Extremum seeking control (ESC) is a model-free optimization technique that uses output measurements to optimize an unknown cost function \cite{KRSTIC2000595,ariyur-2003,TAN2006889,Scheinker2024Survey}. Optimization updates are generated directly from measurements of the cost function without requiring an explicit model of the objective, and thus the explicit derivative evaluation involving the above-mentioned differentiation can be avoided. For example, the averaged dynamics, i.e., the integration over time or iterations, of classical ESC approximate those of gradient descent (GD), resulting in gradient-descent-like optimization behavior without explicit gradient evaluation \cite{TAN2006889}. Motivated by this property, this study investigates the use of ESC as an alternative optimization framework for parametric image registration. 

To assess the feasibility of this approach, we first compare classical ESC and GD for affine image registration using simulated ground-truth motions derived from a beating \emph{ex vivo} porcine heart dataset. Because the averaged dynamics of classical ESC approximate those of GD, this comparison provides a principled framework for assessing the computational efficiency of obtaining optimization updates through ESC relative to explicit derivative evaluation. Building on these results, classical ESC is subsequently employed in both the affine and B-spline stages of a ULM motion-correction pipeline, where affine registration compensates for global tissue motion and B-spline registration corrects residual local deformation. The proposed framework is then applied to super-resolution imaging of a beating \emph{ex vivo} porcine heart to evaluate its effectiveness for motion correction in ULM.

\section{Theory}
In this section, affine image registration is used as an illustrative example to highlight the differences between GD and classical ESC \cite{TAN2006889} in generating optimization updates. While GD relies on automatic differentiation using the chain rule, classical ESC generates update directions by integrating perturbed and demodulated registration costs.

\subsection{Cost Function in Image Registration}

In image registration, a parameterized spatial transformation $T(\bm{x},\bm{\theta})$ is applied to the moving image $I_{\mathrm{mov}}$ to achieve spatial alignment with the reference image $I_{\mathrm{ref}}$, where $\bm{\theta}$ denotes the vector of motion parameters to be estimated. The parameter vector $\bm{\theta}$ is obtained by minimizing a cost function (or maximizing an image-similarity metric) that quantifies the discrepancy between the two images. In this study, the mean squared error (MSE) between image intensities is adopted as the registration cost function:

\begin{equation}
\mathrm{E}(\bm{\theta} ) = \frac{1}{2N_{pixel}}{\sum\limits_x {\left[ {I_\mathrm{ref}(\bm{x}) - {I_\mathrm{mov}}({\mathop{\rm T}\nolimits} (\bm{x},\bm{\theta} ))} \right]} ^2},
\label{eq:MSE_cost}
\end{equation}
where $\bm{x}$ denotes the pixel coordinate and $N_{pixel}$ is the number of pixels. As an example, affine image registration employs a seven-parameter affine transformation model, in which the linear transformation is decomposed into physically interpretable rotation, scaling, and shear components, together with translational motion. The resulting transformation provides a compact representation of global tissue motion and is therefore commonly used as the first stage of motion correction in ULM. For a 2D pixel coordinate $\bm{x}=[x,y]^\mathrm{T}$ in the image plane, the transformation is defined as:
\begin{equation}
\begin{aligned}
\mathrm{T}(\bm{x},\bm{\theta}) =
&\underbrace{
\begin{bmatrix}
\cos\theta & -\sin\theta \\
\sin\theta & \cos\theta
\end{bmatrix}
}_{\mathrm{R}(\phi): \text{Rotation}}
\underbrace{
\begin{bmatrix}
s_x & 0 \\
0 & s_y
\end{bmatrix}
}_{\mathrm{S}(s_x,s_y): \text{Scaling}}
\\[6pt]
&\underbrace{
\begin{bmatrix}
1 & sh_x \\
sh_y & 1
\end{bmatrix}
}_{\mathrm{Sh}(sh_x,sh_y): \text{Shear}}
\, \bm{x}
+
\underbrace{
\begin{bmatrix}
t_x \\
t_y
\end{bmatrix}
}_{\mathrm{T}(t_x,t_y):\text{Translation}}.
\end{aligned}
\label{eq:affine_transformation_matrix}
\end{equation}
The corresponding 7-dimensional parameter vector is:
\begin{equation}
\bm{\theta}  = {\left[ {{t_x},{t_y},\gamma ,{s_x},{s_y},s{h_x},s{h_y}} \right]^{\mathop{\rm T}\nolimits} }.
\label{eq:affine_transformation_array}
\end{equation}
Note that the order of translation, rotation, scaling and shear in Eq. \eqref{eq:affine_transformation_matrix} and  \eqref{eq:affine_transformation_array} needs to be fixed during the optimization, but may vary depending on the implementation \cite{yoo2002ITK,jenkinson2002improved}.

\subsection{Gradient Descent for Parametric Image Registration}\label{sub_section_GD}

In parametric image registration, the gradient descent (GD) method updates the motion parameters along the negative gradient direction of the cost function in Eq.~\eqref{eq:MSE_cost}. At each iteration, the parameter vector is updated from the $(k-1)^{\mathrm{th}}$ iteration to the $k^{\mathrm{th}}$ iteration according to
\begin{equation}
{\bm{\theta} ^{k}} = {\bm{\theta} ^{k-1}} - \rm K_{GD}\nabla \mathrm{E}({\bm{\theta}}),
\label{eq:GD_update}
\end{equation}
where $\rm K_{GD}$ is the step length and serves as a tuning hyperparameter.

%

The gradient of Eq.~\eqref{eq:MSE_cost} with respect to (w.r.t) the motion parameters $\bm{\theta}$, defined in (\ref{eq:affine_transformation_matrix}), is obtained through automatic differentiation using the chain rule:

\begin{align}
\frac{{\partial \mathop{\rm E}}}{{\partial {\theta _i}}} = \frac{1}{N_{pixel}} \sum\limits_x  & {term_1 \times term_2  \times term_3 } \label{eq:chain_rule}\\
term_1 & = I_\mathrm{mov}({\mathop{\rm T}} (\bm{x},\bm{\theta} ))-I_\mathrm{ref}(\bm{x})  \tag{\ref{eq:chain_rule}{a}} \label{eq:chain_rule_term1}\\
term_2 & = \nabla I_\mathrm{mov}({\mathop{\rm T}} (\bm{x},\bm{\theta} )) \tag{\ref{eq:chain_rule}{b}} \label{eq:chain_rule_term2}\\
term_3 & = \frac{{\partial {{\mathop{\rm T}} (\bm{x},\bm{\theta} )} }}{{\partial \bm{\theta}}} \tag{\ref{eq:chain_rule}{c}}. \label{eq:chain_rule_term3} 
\end{align}
These formulations show that gradient evaluation requires the computation of pixel-wise intensity differences in Eq.~\eqref{eq:chain_rule_term1}, image-intensity gradients with respect to pixel displacement in Eq.~\eqref{eq:chain_rule_term2}, and pixel-displacement gradients with respect to the motion parameters in Eq.~\eqref{eq:chain_rule_term3}, followed by accumulation over all pixels as indicated in Eq.~\eqref{eq:chain_rule}. As image resolution and the number of motion parameters increase, the computational burden of gradient evaluation grows significantly, resulting in increased computational cost at each iteration.





\subsection{Extremum Seeking Control for Parametric Image Registration}\label{sub_section_ESC}

\begin{figure}[htb!]
    \centering
\includegraphics[width=7.5 cm]{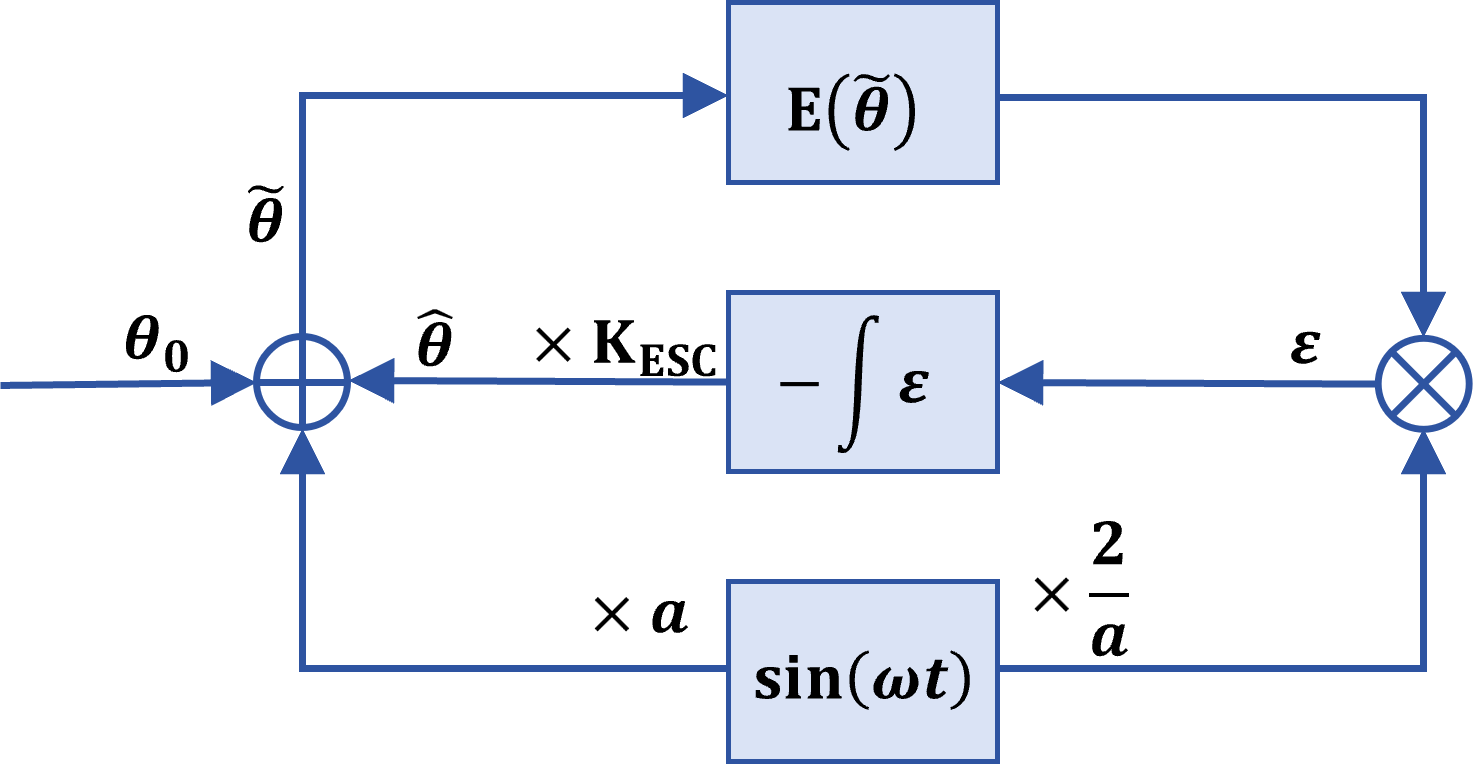}
    \caption{A basic ESC framework. ESC perturbs the parameters $\bm \theta$ with sinusoidal waves $a\sin({\bm \omega} t)$, demodulates the caused variations in the cost function ${\mathop{\rm E}\nolimits}$  with sinusoidal waves $2\sin({\bm \omega} t)/a$, and integrates demodulated variations $\varepsilon$ to update the parameters $\bm \theta$.}
    \label{fig:ESC_diagram}
\end{figure}

The extremum seeking control (ESC) method provides an alternative approach for solving optimization problems without requiring explicit gradient evaluation. ESC can be applied to both continuous-time \cite{TAN2006889} and discrete-time optimization problems\cite{Teel2001ESC}. In this paper, we employ the classical ESC algorithm, illustrated in Fig.~\ref{fig:ESC_diagram}, whose averaged dynamics approximate those of the GD method introduced in the previous subsection.

The classical ESC algorithm operates in continuous time. Small-amplitude periodic perturbations are introduced into the optimization parameters, producing corresponding variations in the cost function. These cost-function measurements are then processed through temporal integration to generate parameter updates. The resulting update law is given by
\begin{equation}
\widetilde{\bm{\theta}} (t) = {\bm{\theta} _0}+ \hat {\bm{\theta}} (t) + a\sin (\bm{\omega} t)
\label{eq:ESC_continuous_time_modulate}
\end{equation}
\begin{equation}
\dot{\hat{\bm{\theta}}}(t) =  -\frac{\rm 2K_{ESC}}{a}{\mathop{\rm E}\nolimits} (\widetilde{\bm{\theta}} (t) )\sin (\bm{\omega} t)
\label{eq:ESC_continuous_time_update}
\end{equation}
where $a$ is the perturbation amplitude, $\bm{\omega}$ is a column vector of perturbation frequencies associated with each parameter, $\bm{\theta}_0$ is the initial parameter vector, $\hat{\bm{\theta}}(t)$ denotes the slowly varying parameter estimate generated by the ESC process, and $\rm K_{ESC}$ is the gain used for ESC algorithm.

Although the convergence properties of classical ESC have been rigorously established in \cite[Corollary 1]{TAN2006889}, a brief derivation is presented here to illustrate how the algorithm extracts descent information. To this end, the cost function is expanded using a first-order Taylor series around the current parameter value $\bm{\theta}_\mathrm{c} = {\bm{\theta}_0}+\hat{\bm{\theta}}(t_c)$:
\begin{equation}
{\mathop{\rm E}\nolimits} (\widetilde{\bm{\theta}} (t) ) \approx {\mathop{\rm E}\nolimits} ({\bm{\theta} _{\mathop{\rm c}\nolimits} )} + a\nabla {\mathop{\rm E}\nolimits} {\left( {{\bm{\theta} _{\mathop{\rm c}\nolimits} }} \right)^{\mathop{\rm T}\nolimits} }\sin (\bm{\omega} t).
\end{equation}

Multiplying the above expression by the demodulation signal $\frac{2}{a}\sin(\bm{\omega}t)$ yields the demodulated signal $\bm{\varepsilon}=\frac{2}{a}{\mathop{\rm E}\nolimits}(\widetilde{\bm{\theta}}(t))\sin(\bm{\omega}t)$, which has the same dimension as the parameter vector $\bm{\theta}$. The $j^{\mathrm{th}}$ component corresponding to parameter $\theta_j$ is given by
%
%
%
%
\begin{equation}
{\varepsilon_{j}}  \approx \frac{2}{a}{\mathop{\rm E}\nolimits} ({\bm{\theta} _\mathrm{c}})\sin ({\omega}_j t) + 2\sum\limits_{i = 1}^{\mathop{\rm n}\nolimits}  \frac{{\partial {\mathop{\rm E}\nolimits} \left( {{\bm{\theta} _\mathrm{c}}} \right)}}{{\partial {\theta _i}}}\sin ({\omega _i}t)\sin ({\omega}_j t) , 
\end{equation}
where ${\omega}_i$ and ${\omega}_j$ are perturbation frequencies contained in $\bm{\omega}$. By selecting the perturbation frequencies sufficiently large, the parameter-update dynamics evolve on a slower time scale than the sinusoidal perturbations. Consequently, averaging techniques \cite{khalil-2002} can be applied over one or more perturbation periods.  
The first term is a sinusoidal oscillation at the demodulation frequency ${\omega}_j$. Applying time averaging over one or more periods yields
\begin{equation}
\left\langle {\frac{2}{a}{\mathop{\rm E}\nolimits} ({\bm{\theta}_c } )\sin ({\omega}_j t)} \right\rangle  = 0.
\end{equation}
For the second term, the time averaging yields:
%
%
\begin{equation}
\left\langle {\sin \left( {{\omega _i}t} \right)\sin \left( {{\omega _j}t} \right)} \right\rangle  = \left\{ {\begin{array}{*{20}{c}}
{0,{\rm{  }}i \ne j}\\
{{\rm{ }}\frac{1}{2},{\rm{  }}i = j{\rm{ }}}, 
\label{eq:ESC_orthogonality_property}
\end{array}} \right.
\end{equation}
where the cross-frequency terms vanish because the perturbation frequencies are mutually distinct and sufficiently separated. Consequently, the time-averaged demodulated signal becomes proportional to the corresponding gradient component:
\begin{equation}
\left\langle  \varepsilon_j  \right\rangle    \propto  \frac{{\partial {\mathop{\rm E}\nolimits} \left( {{\bm{\theta} _\mathrm{c}}} \right)}}{\partial\theta_j}.
\end{equation}
This shows that  ESC can estimate or approximate the descent direction mainly with a computation of the similarity
metric and without explicit computation of gradients.

 For the discrete iterations in parameteric image registration, Eq. \eqref{eq:ESC_continuous_time_modulate}   and \eqref{eq:ESC_continuous_time_update}   can be rewritten as following
\begin{equation}
\widetilde{\bm{\theta}}^k = {\bm{\theta}}^0 + {\bm{\theta}}^{k-1}  + a\sin (\bm{\omega} k),
\label{eq:ESC_discrete_time_modulate}
\end{equation}
\begin{equation}
\bm{\theta}^{k} = \bm{\theta}^{k-1} -\frac{\rm {2K_{ESC}}}{a}{\mathop{\rm E}\nolimits} (\widetilde{\bm{\theta}}^{k})\sin (\bm{\omega} k).
\label{eq:ESC_discrete_time_update}
\end{equation}

It is worthwhile to note that ESC is not the only model-free optimization technique. Finite-difference methods, for instance, have been widely used to estimate gradients from cost-function evaluations alone, has been used in parametric image registration\cite{klein2007evaluation}. Compared with forward finite differences, which require n+1 cost-function evaluations to estimate the gradient of an n-parameter problem, classical ESC employs simultaneous orthogonal perturbations and recovers gradient information through demodulation and temporal averaging, requiring only a single cost-function evaluation per iteration.

\subsection{GD versus ESC for Parametric Image Registration}

To quantify the computational advantage of ESC (\ref{eq:ESC_discrete_time_update}) over GD in \eqref{eq:chain_rule}, Table~I compares the computational cost per iteration of the two methods using affine image registration with image dimensions of 371 $\times$ 371 pixels as an example. The results show that GD implemented with automatic differentiation requires 27,803,533 operations per iteration, whereas ESC requires only 7,845,630, making the computational cost of GD approximately 3.54 times that of ESC.

\begin{table*}[ht]
\centering
  \begin{threeparttable}
\caption{Comparison of computational complexity per iteration between gradient-based and ESC methods}
\label{tab:complexity}
\tiny
\begin{tabular}{cccccc}
\toprule
Module &
Function Description &
Operation Type &
GD &
ESC \\
\midrule

\multirow{2}{*}{A--Eq. \eqref{eq:MSE_cost} \eqref{eq:affine_transformation_matrix}}
&
\multirow{2}{*}{\makecell[c]{Common Module$^{a}$}}
& $\times$
& $20+N_{\mathrm{pixel}}\times21$ $(2890481)$
& $20+N_{\mathrm{pixel}}\times21$ $(2890481)$
\\

&& $+$
& $3+N_{\mathrm{pixel}}\times36$ $(4955079)$
& $3+N_{\mathrm{pixel}}\times36$ $(4955079)$
\\

\multirow{2}{*}{B--Eq. \eqref{eq:chain_rule_term1}}
&
\multirow{2}{*}{Intensity Difference}
& $\times$
& $0$
& ---
\\

&& $+$
& $N_{\mathrm{pixel}}$ $(137641)$
& ---
\\

\multirow{2}{*}{C--Eq. \eqref{eq:chain_rule_term2}}
&
\multirow{2}{*}{\makecell[c]{Gradient of Image Intensity \\ w.r.t. Pixel Displacement}}
& $\times$
& $0$
& ---
\\

&& $+$
& $2N_{\mathrm{pixel}}$ $(275282)$
& ---
\\

\multirow{2}{*}{D--Eq. \eqref{eq:chain_rule_term3}}
&
\multirow{2}{*}{\makecell[c]{Gradient of  Pixel Displacement \\ w.r.t. Parameter}}
& $\times$
& $101N_{\mathrm{pixel}}$ $(13901741)$
& ---
\\

&& $+$
& $34N_{\mathrm{pixel}}$ $(4679794)$
& ---
\\

\multirow{2}{*}{E--Eq. \eqref{eq:chain_rule}}
&
\multirow{2}{*}{Average Across Pixels}
& $\times$
& $3N_{\theta}$ $(21)$
& ---
\\

&& $+$
& $N_{\theta}(N_{\mathrm{pixel}}-1)$ $(963480)$
& ---
\\

\multirow{2}{*}{F--Eq. \eqref{eq:GD_update}}
&
\multirow{2}{*}{Parameter Update (GD)}
& $\times$
& $N_{\theta}$ $(7)$
& ---
\\

&& $+$
& $N_{\theta}$ $(7)$
& ---
\\

\multirow{2}{*}{G--Eq.\eqref{eq:ESC_discrete_time_modulate}\eqref{eq:ESC_discrete_time_update}}
&
\multirow{2}{*}
{\makecell[c]{Parameter Modulate, Cost demodulated, \\ Parameter Update (ESC)}}

& $\times$
& ---
& $7N_{\theta}$ $(49)$
\\

&& $+$
& ---
& $3N_{\theta}$ $(21)$
\\

All
&
---
&
$\times$
&
16792243
&
2890530
\\

&
&
$+$
&
11011290
&
4955100
\\


Total
&
---
&
---
&
27803533
&
7845630
\\

\bottomrule
\end{tabular}
  
\begin{tablenotes}
\footnotesize
    \item[a] Computation for deforming images by affine transformation and bilinear interpolation, and calculating the similarity metric.
    \end{tablenotes}
  
\end{threeparttable}
    
\end{table*}



\begin{figure*}[htb!]
    \centering
\includegraphics[width=16cm]{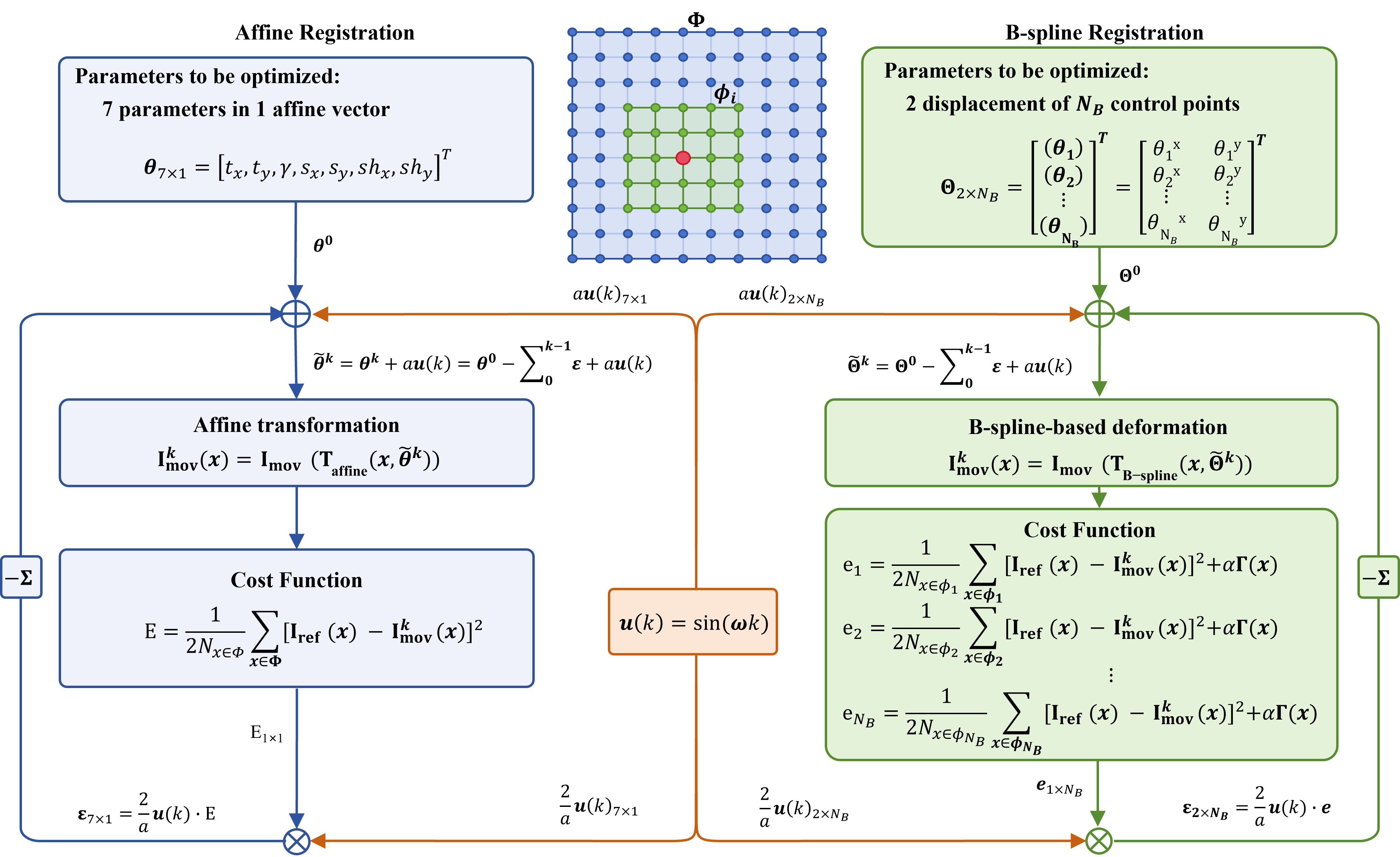}
    \caption{Diagram of the two-stage image registration method based on ESC. Transformation and deformation is only implemented once in each iteration after perturbing the parameters. The cost function is calculated once across the whole image for parameters in affine registration or across a sub-region for parameters of the corresponding control point. $\mathbf{\Gamma}(\bm x)$ is the 2-D bending energy of a thin-plate of metal function, and $\alpha$ is the regularization weight. $\rm E$ is the cost for the global registration, and $\rm e$ is the cost for the local registration.}
    \label{fig:image_registration_diagram}
\end{figure*}

\section{Method}

This section presents the implementation of ESC for two-stage image registration, consisting of affine registration followed by B-spline-based non-rigid registration. It then compares GD and ESC, introduced in Sections \ref{sub_section_GD} and \ref{sub_section_ESC}, respectively, in an affine registration task through simulation studies, where the source images and motion fields are derived from an \emph{ex vivo} pig heart. Finally, the feasibility of the proposed ESC-based two-stage image registration framework is demonstrated using ULM imaging data acquired from the same \emph{ex vivo} pig heart.

\subsection{ULM Reconstruction with ESC-Based Motion Correction}


The implementation of affine and B-spline registration is illustrated in Fig. \ref{fig:image_registration_diagram}. The affine transformation is defined by Eq. \eqref{eq:affine_transformation_matrix}, while the B-spline deformation model follows \cite{rueckert2002nonrigid}. The Mean Squared Error (MSE) is adopted as the cost function. In both registration methods, ESC applies periodic perturbations to the motion parameters, updates the moving image through affine transformation or B-spline deformation, evaluates the cost using the reference image and the transformed image, demodulates the cost using the perturbation signals, and integrates the demodulated signals to estimate the motion parameters.

For affine registration, seven distinct perturbation frequencies $\bm{u}_{7\times1}$ are assigned to the seven affine parameters in Eq. \eqref{eq:affine_transformation_array}. The cost is evaluated once per iteration over the entire image region ($\Phi$). The resulting cost is then demodulated by each perturbation signal to obtain the update direction for the corresponding affine parameter.

For B-spline registration, a minimum two-dimensional B-spline grid consisting of 4 $\times$ 4 control points is adopted. Since each control point contains lateral and axial displacement components, 4$\times$ 4$\times$ 2 = 32 distinct perturbation frequencies are required. These frequencies are repeatedly assigned across the entire B-spline grid such that the 16 control points influencing any pixel are associated with mutually distinct perturbation frequencies, satisfying the orthogonality requirement of ESC in Eq. \eqref{eq:ESC_orthogonality_property}.
For the $i^{\mathrm{th}}$ control point, the cost $e_i$ for local registration is evaluated over a local region ($\phi_i$) extending two grid spacings from the control point, corresponding to the area whose deformation is influenced by that control point.  The local cost is demodulated using the perturbation signals associated with the lateral and axial displacement parameters of the control point. The entire image is deformed only once per iteration.

\subsection{Ultrasound Data}

A 5-second ultrasound dataset comprising 1500 frames acquired from an \emph{ex vivo} porcine heart was used for evaluation and demonstration. The heart was explanted from a large white female pig (65--75 kg, 4--5 months old) and maintained using a Langendorff \emph{ex vivo} perfusion model. Sonovue microbubbles (Bracco, Milan, Italy) were infused into the heart using a syringe pump (Harvard Apparatus, Holliston, US) at a rate of 5 mL/min.
The animal study was reviewed and approved by the Royal Veterinary College Animal and Ethical Review Board and was conducted in accordance with the relevant ethical regulations, including European Commission 2010, the Animal Welfare Act 2006, and the Welfare of Farmed Animals (England) Regulations 2007.

Ultrasound data were acquired using a research ultrasound system (Vantage 256, Verasonics, US) and a 1.5D phased-array probe (M5ScD, GE Healthcare, US) with a central frequency of 2.84 MHz, a pitch of 0.27 mm, and 80 $\times$ 3 elements, consisting of one central row and two side rows of 80 half-height elements. Ultrafast imaging was performed using diverging-wave transmissions steered at six angles, with a transmit center frequency of 2.4 MHz and an angular field of view of 53$^\circ$.

Amplitude modulation (AM) imaging was implemented using full-, half-, and half-interleaved-aperture transmissions at each steering angle to separate microbubble signals from tissue signals. The effective frame rate after angle compounding and AM processing was 305$~\mathrm{Hz}$. Further details of the experimental setup and imaging sequence are provided in our previous work \cite{yan2024transthoracic}.
B-mode and CEUS images were reconstructed onto a 371 $\times$ 371 pixel grid with a pixel size of 0.27 $~\mathrm{mm}$ using the Delay-and-Sum beamformer \cite{perrot2021so} and the Coherence-to-Variance beamformer \cite{yan2023fast}, respectively.

\subsection{Simulation-Based Comparison of GD and ESC}

\begin{figure*}[!htb]
    \centering
\includegraphics[width=18 cm]{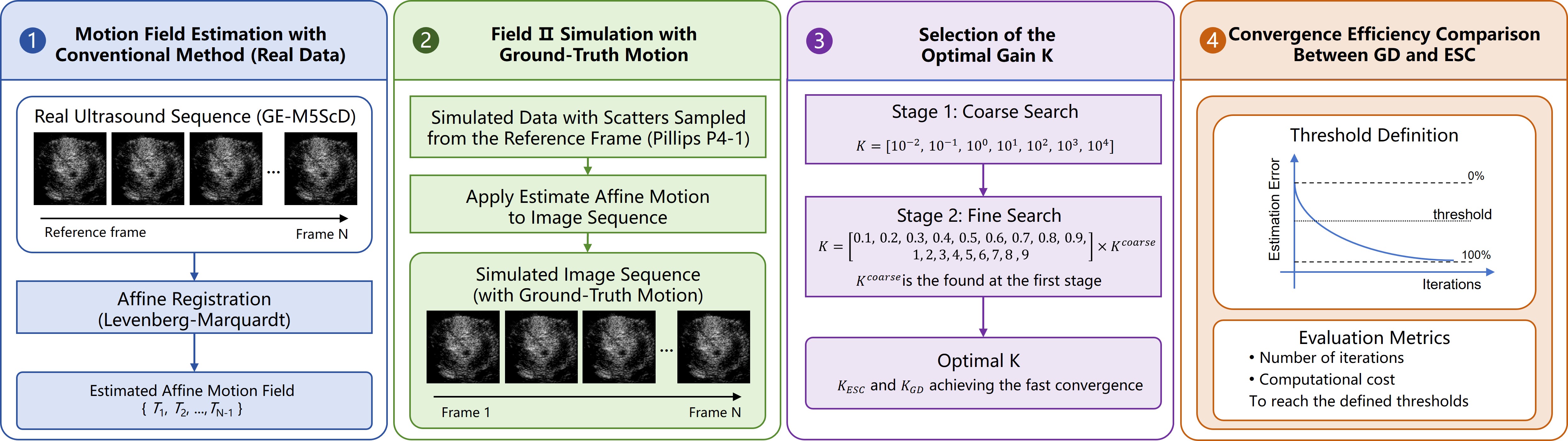}
    \caption{Diagram of the comparison between ESC and GD for affine image registration. Affine motions are estimated by another algorithm from acquired data and applied to a image simulated by Field II with another probe. ESC and GD are compared at the aspects of iteration convergence speed and computation cost after searching their optimal gain.}
    \label{fig:comparison_diagram}
\end{figure*}

GD and ESC are compared using simulated image sequences with known ground-truth affine motion. The ground-truth motion fields are estimated from B-mode images acquired from the \emph{ex vivo} porcine heart dataset and applied to ultrasound images simulated using Field II, where scatterers were sampled from the experimental images. To ensure a fair comparison, the gain parameter of each method is optimized independently to achieve its best convergence performance. The overall comparison framework is illustrated in Fig. \ref{fig:comparison_diagram}.

\subsubsection{Generation of Motion Fields from Experimental Data}

A reference frame was selected as the B-mode image with the highest structural similarity, calculated using MATLAB's \emph{ssim} function, to the image obtained by temporally averaging the entire sequence. Affine motion parameters were estimated between the reference frame and each subsequent frame by minimizing the MSE using the Levenberg--Marquardt algorithm. Only non-zero affine parameter sets were retained for generating motion fields consisting of pixel-wise displacement vectors.

\subsubsection{Field II-Based Ultrasound Simulation}

A total of 10,000 scatterers were randomly sampled from the reference image, with pixel locations and intensities assigned as scatterer positions and amplitudes, respectively. Ultrasound data were simulated using a Philips P4-1 probe with a center frequency of 2.5 MHz, a pitch of 0.295 mm, and 96 elements. Diverging-wave transmissions were simulated with a field of view of 120$^\circ$ and seven steering angles ranging from -15$^\circ$ to +15$^\circ$. The simulated radiofrequency data were beamformed using Delay-and-Sum (DAS) onto the same image grid as the reference image. The motion fields estimated from the experimental data were then applied to the reference image to generate image sequences with known ground-truth affine motion.

\subsubsection{Selection of the Optimal Gain $K_{ESC}$ and $K_{GD}$}

To ensure a fair comparison between GD and ESC, the gain parameters $K_{GD}$ and $K_{ESC}$ are independently optimized to achieve the fastest convergence while maintaining stability. For each method, the optimal gain is selected by minimizing the normalized global average registration Error (nGARE), defined as 
\begin{equation}
{\rm{nGARE}_{K}} = \frac{1}{N_fN_{k}}\sum\limits_{f=1}^{N_f} {\sum\limits_{k=1}^{N_k} \frac{D_{k,f}}{D_{0,f}}},  \
\label{eq:GARE_definition}
\end{equation}
where $N_f$ is the number of frames used for evaluation, $N_k$ is the number of iterations, $D_{k,f}$ is the mean displacement estimation error over all pixels at the $k^{\mathrm{th}}$ iteration for the $f^{\mathrm{th}}$ frame, and $D_{0,f}$ is the corresponding initial displacement error. Ten frames with initial motion magnitudes ranging from large to small were empirically selected from the simulated image sequence for evaluation, and $N_k$ was set to 1000. To penalize unstable or divergent behavior, all values of $D_{k,f}$ for a frame are assigned as $10D_{0,f}$ if either of the following conditions is satisfied: (i) more than half of the iterations produce errors larger than $D_{0,f}$, or (ii) for GD, the final error satisfies $D_{1000,f}>\min(D_{k,f})$. The optimal gain is determined using a two-stage search procedure. In the first stage, a coarse search is conducted over a broad range of candidate values, $K_*^{coarse} = \left[ 10^{-2},\, 10^{-1},\, 10^{0},\, 10^{1},\, 10^{2},\, 10^{3},\, 10^{4} \right]$, where $*\in \left\{GD,ESC\right\}$. 
In the second stage, a finer search is performed around the best value identified during the coarse search. Additional candidate gains are generated by linearly interpolating within the neighboring interval surrounding the selected $K_{*}^{\mathrm{coarse}}$. For example, if the optimal coarse-search result is $K_*^{\mathrm{coarse}}=10$, the fine-search candidates are $K^{fine} = \left[ 1,\, 2,\, 3,\,  4,\,  5,\,  6,\,  7,\,  8,\,  9,\,  10,\,  20,\,  30,\,  40,\,  50,\,  60,\,  70,\,  80,\,  90\right]$ if the best $K_{*}^{coarse}$=10.

\subsubsection{Performance Evaluation}

Affine registration using GD and ESC with their respective optimal gains is applied to the complete simulated image sequence. For each image frame, the displacement estimation error is recorded throughout the optimization process.

To compare convergence performance, the initial displacement estimation error of each frame is defined as the upper bound. The larger of the final errors obtained by GD and ESC is defined as the lower bound. Three intermediate thresholds corresponding to $25\%$, $50\%$, and $75\%$ error reduction between the upper and lower bounds are then determined.

Due to the oscillatory behavior of ESC, a threshold is considered reached only when all subsequent errors remain below that threshold. The iteration number at which this condition is first satisfied is recorded as the convergence iteration for the corresponding threshold.

Computational efficiency is evaluated using both the number of iterations required to reach each threshold and the corresponding computational cost, calculated as the number of iterations multiplied by the computational cost per iteration. Statistical comparisons between GD and ESC are performed at each threshold using the Wilcoxon signed-rank test implemented by MATLAB's \emph{signrank} function. The rank-biserial correlation coefficient $r_{rb}$ is reported as the effect-size measure.

\subsection{ULM Reconstruction with ESC-Based Motion Correction}

Tissue motion is estimated from B-mode images using the proposed two-stage ESC registration framework shown in Fig. \ref{fig:image_registration_diagram}. The estimated motion is then used to correct the corresponding CEUS frames. After noise suppression by thresholding, microbubbles (MBs) are isolated using normalized cross-correlation between the CEUS images and an estimated point spread function, followed by centroid-based localization. The localized MBs are subsequently tracked using the feature-motion-model framework \cite{yan2022super,yan2024transthoracic}.
ULM images are reconstructed by accumulating the tracked MB trajectories onto a grid with a pixel size of 27 ${\rm \mu}m$ and subsequently smoothing the image using a Gaussian kernel with a full width at half maximum (FWHM) equal to one quarter of the ultrasound wavelength. The localization, tracking, and reconstruction methods are publicly available on GitHub (\url{https://github.com/JipengYan1995/SRUSSoftware}).

For comparison, a ULM image is also reconstructed from the CEUS sequence without motion correction. Image resolution is quantified using Fourier ring correlation (FRC) analysis. To avoid bias arising from regions with low microbubble saturation \cite{hingot2021measuring}, FRC is evaluated within a region of interest containing a highly saturated vessel rather than over the entire image. The analysis is performed by dividing the tracked MB trajectories into two independent groups.

\section{Results}

\begin{figure}[h]
    \centering
\includegraphics[width=7.5 cm]{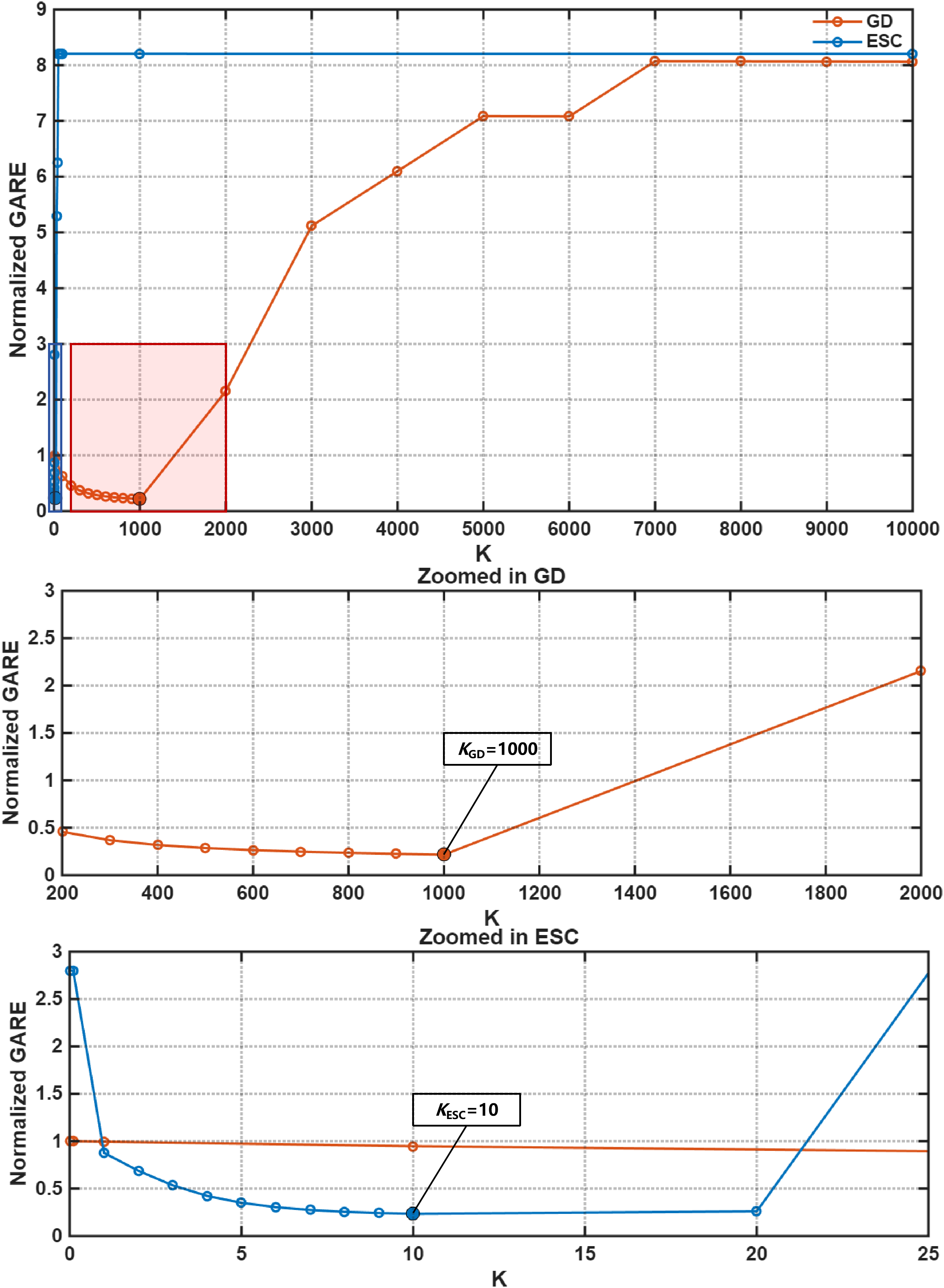}
    \caption{Normalized GARE versus gain $K$ in the iterations for GD and ESC. The optimal gains $K$ are found by the coarse and fine search as 1000 and 10 for GD and ESC respectively. }
    \label{fig:Error_vs_K}
\end{figure}

The influence of the gain parameter $K$ on GD and ESC is shown in Fig. \ref{fig:Error_vs_K}. The performance of both methods is highly dependent on the choice of $K$. Since $D_{k,f}/D_{0,f}$ in Eq. \eqref{eq:GARE_definition} is expected to decrease from 1 to 0 as the optimization progresses, normalized nGARE values greater than 1 indicate unstable or divergent behavior. Owing to the penalty term introduced in Eq. \eqref{eq:GARE_definition}, excessively small gains may also result in nGARE values greater than 1 due to slow convergence, while oscillatory behavior in ESC can similarly increase nGARE when $D_{k,f}$ temporarily exceeds $D_{0,f}$.

Using the two-stage search procedure described in Fig. \ref{fig:comparison_diagram}, the optimal gains were determined as $K_{ESC}=10$ and $K_{GD}=1000$. These values were used for all subsequent analyses.

\begin{figure}[!htb]
    \centering
\includegraphics[width=8 cm]{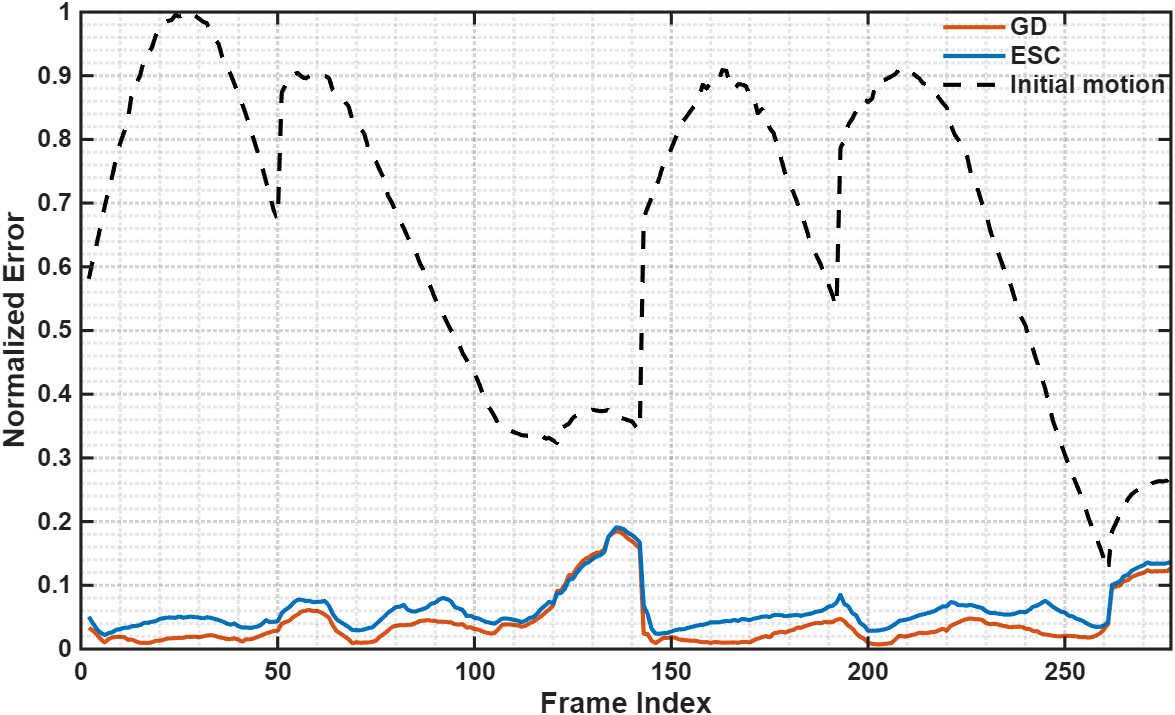}
    \caption{Normalized motion estimation error after optimization for all the simulated frames. The initial motion is calculated by the mean displacement distance in the ground truth motion field of each frame and normalized by the maximum initial motion in the sequence. The error for GD or ESC is calculated as the mean distance between the simulated and estimated motion field, and normalized by the maximum initial motion.}
    \label{fig:final_error_across_frames}
\end{figure}

\begin{figure}[ht]
    \centering
\includegraphics[width=7.5 cm]{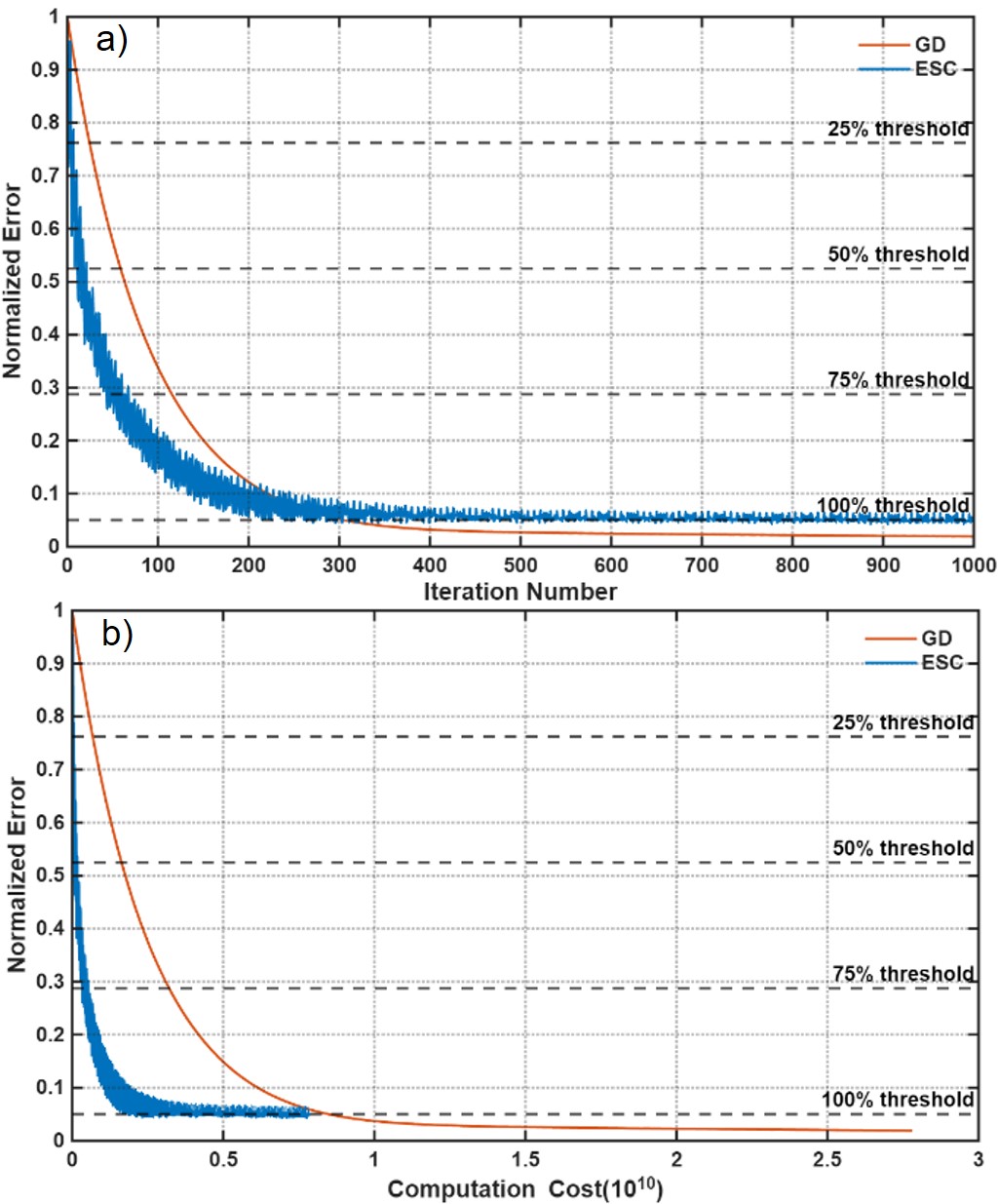}
    \caption{Convergence curves of two methods for the $27^{th}$ frame with the iteration number as the horizontal axis a) and the computation cost as the horizontal axis b). }
    \label{fig:Threshold_definition}
\end{figure}

\begin{figure}[ht]
    \centering
\includegraphics[width=7.5 cm]{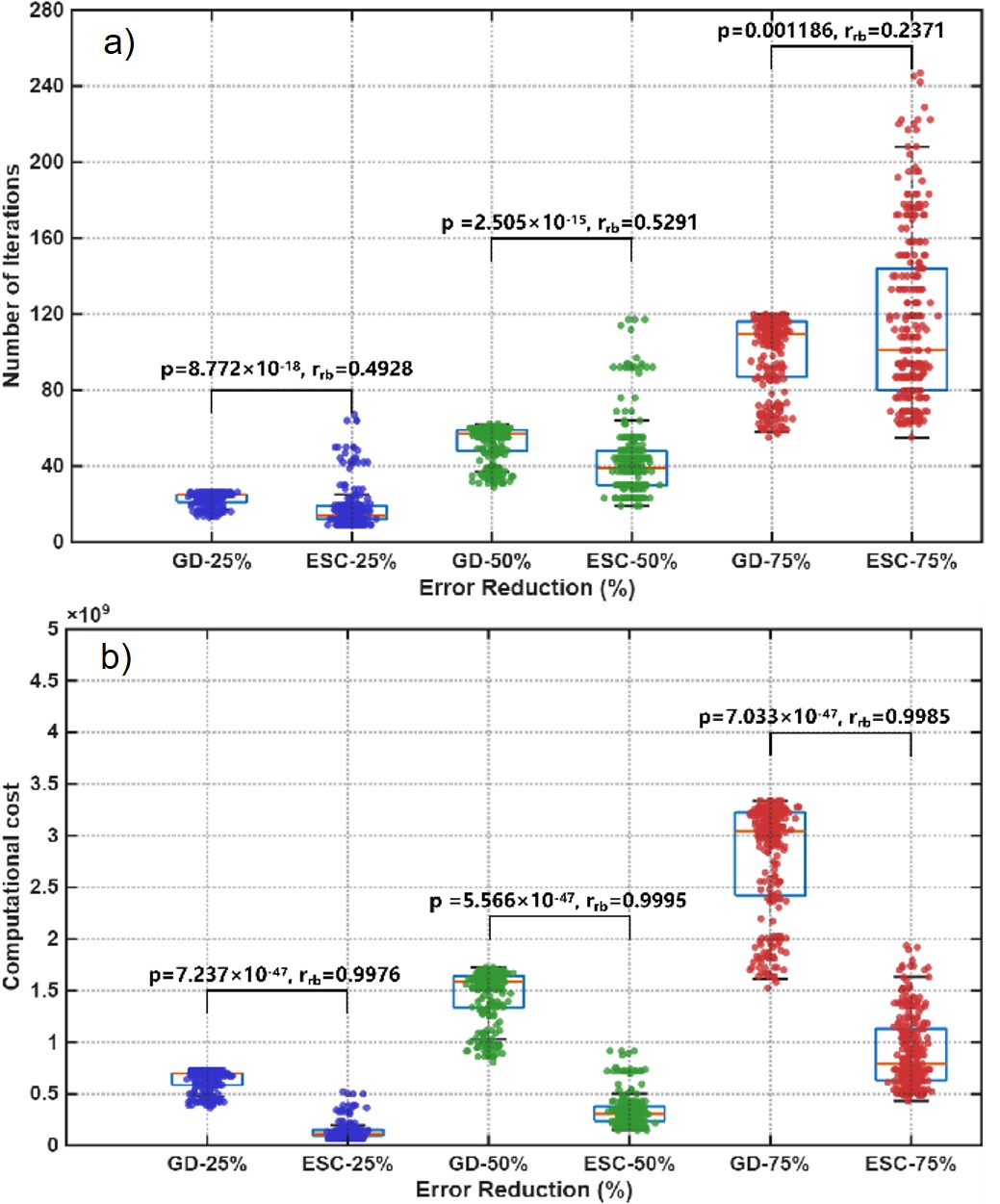}
    \caption{Computation efficiency analysis. a) Box plot of iterations required to reach 25$\%$, 50$\%$, and 75$\%$ thresholds. b) Box plot of computational cost required to reach 25$\%$, 50$\%$, and 75$\%$ thresholds.}
    \label{fig:Computation_Efficiency_analysis}
\end{figure}

The equivalence and differences between GD and ESC are illustrated in Figs. \ref{fig:final_error_across_frames}, \ref{fig:Threshold_definition}, and \ref{fig:Computation_Efficiency_analysis}. Fig. \ref{fig:final_error_across_frames} presents the motion estimation errors after 1000 iterations. The mean motion estimation errors were 0.042$\pm$0.040 and 0.062$\pm$0.034 for GD and ESC, respectively, indicating that both methods effectively reduced the motion and achieved comparable registration accuracy.

Fig. \ref{fig:Threshold_definition} illustrates the convergence behavior of GD and ESC for a representative frame in terms of both iteration count and computational cost. Statistical comparisons across all evaluated frames are presented in Fig. \ref{fig:Computation_Efficiency_analysis}. When convergence speed is measured by the number of iterations required to reach each threshold, the Wilcoxon signed-rank test indicates a statistically significant difference between the two methods. However, the corresponding rank-biserial correlation coefficients indicate only a small effect size, and the distributions of the required iterations exhibit substantial overlap.
In contrast, when convergence speed is evaluated using computational cost, the difference between the two methods is statistically significant and associated with a large effect size ($r_{rb}>$0.995) at all thresholds. Namely, the computational costs of ESC are lower than those of GD for nearly all paired comparisons. These results indicate that ESC achieves convergence with substantially lower computational cost while maintaining comparable registration accuracy.

The B-mode sequence, tracked microbubble (MB) trajectories, and the overlap between ULM density maps before and after motion correction are provided in the supplementary video. Fig. \ref{fig:ULM density map} presents the reconstructed ULM images before and after motion correction, together with magnified views of the upper central region. The motion-corrected ULM image exhibits substantially sharper and more clearly defined vascular structures.

\begin{figure}[ht]
    \centering
    \includegraphics[width=7.5 cm]{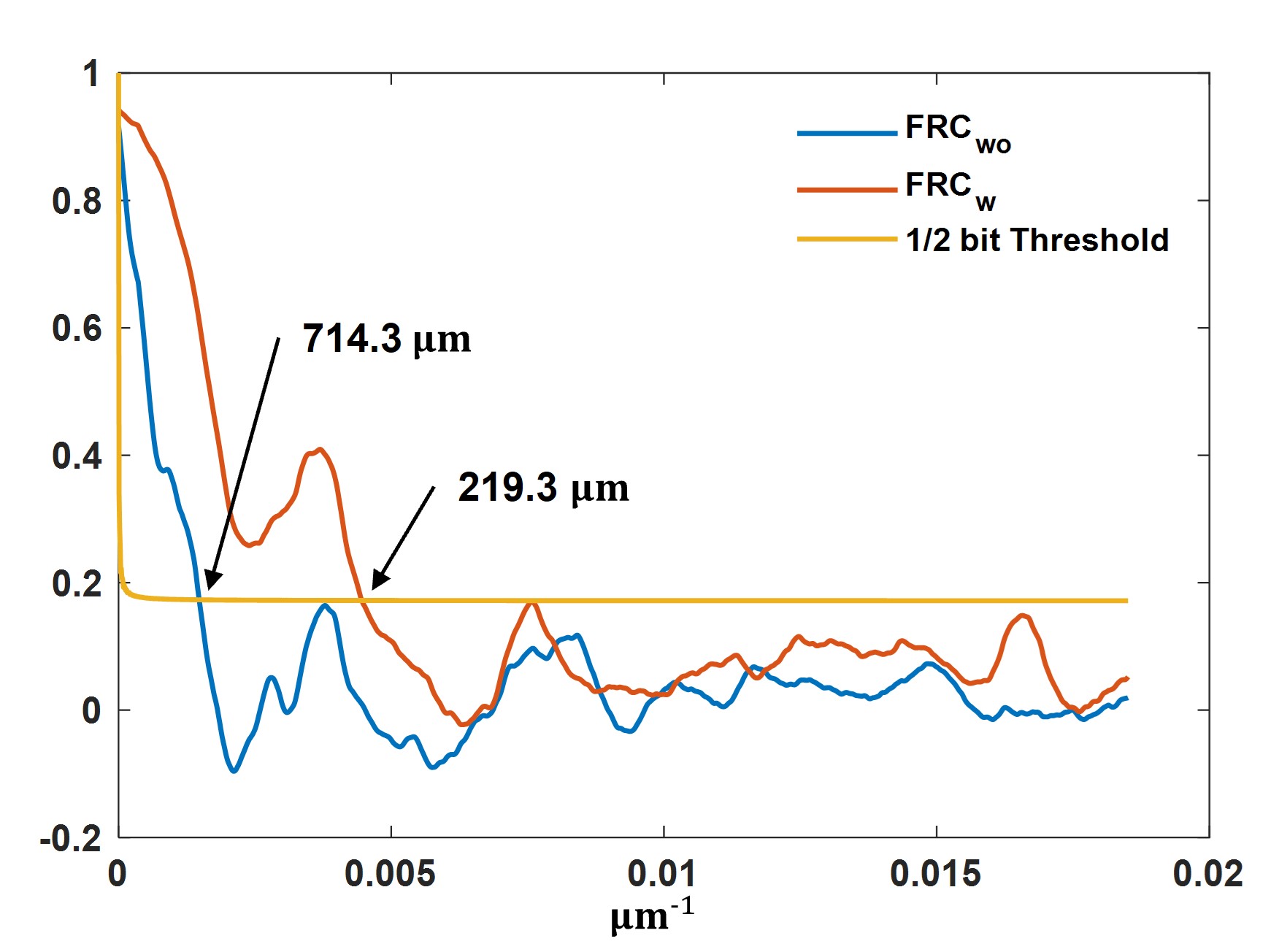}
    \caption{FRC analysis on images without (wo) and with (w) tissue motion correction. Note that the half wavelength of transmitted waves is 1540/2.4/2 = 320.8 $\mu m$. }
    \label{fig:FRC_curves}
\end{figure}

As shown in Fig. \ref{fig:FRC_curves}, Fourier ring correlation (FRC) analysis measured a resolution of 714.3 $\mu m$ for the ULM image reconstructed without motion correction and 219.3 $\mu m$ for the motion-corrected ULM image. This improvement demonstrates that tissue motion correction is essential for achieving super-resolution imaging in ULM and confirms the effectiveness of the proposed two-stage ESC-based registration framework for correcting tissue motion in experimental data.

\begin{figure*}[ht]
    \centering
    \includegraphics[width=17.5 cm]{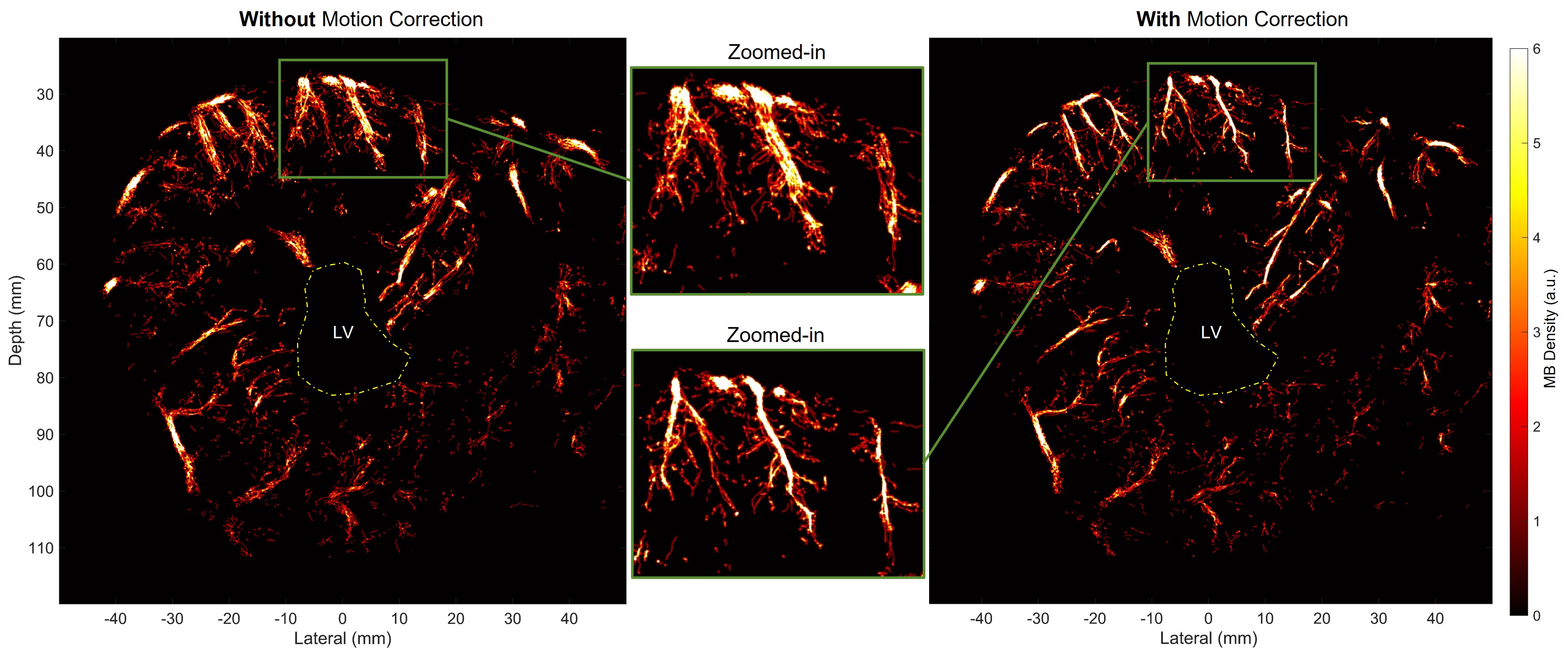}
    \caption{ULM density maps reconstructed from acquired 5-seconds CEUS sequence without and with the two-stage ESC-based motion correction. LV: left ventricle.}
    \label{fig:ULM density map}
\end{figure*}

\section{Discussion}

This study investigated the feasibility of applying ESC to parametric image registration for tissue motion correction in ULM. The results demonstrate that ESC achieves motion correction performance comparable to GD while substantially reducing computational cost. Although the two methods produced similar registration accuracy and required a comparable number of iterations to reach convergence, ESC consistently required less computation to achieve the same level of motion correction.

The computational advantage of ESC originates from its mechanism for obtaining gradient information. Conventional gradient-based methods require explicit gradient computation, whereas ESC estimates the gradient direction implicitly from the response of the cost function to injected dither signals. Consequently, ESC requires only a single cost-function evaluation in each iteration, making its computational cost less sensitive to the dimensionality of the transformation parameters. The results demonstrate that this property enables ESC to achieve comparable registration performance to GD with substantially lower computational cost. 

To ensure a fair comparison, the simplest forms of ESC and GD were adopted. While many variants of both methods exist, the basic formulations require the fewest hyper-parameters and are theoretically comparable. In addition, the ground-truth motion fields were estimated using a separate optimization algorithm, namely Levenberg--Marquardt, and the simulated ultrasound data were generated using a probe different from that used to acquire the experimental data. These measures reduced the possibility of introducing an inherent preference toward either optimization method. As shown in Fig. \ref{fig:Error_vs_K}, the performance of both methods is highly sensitive to the gain parameter, and therefore each method is evaluated using its independently optimized gain. Under these conditions, GD and ESC achieved similar motion estimation accuracy and comparable convergence behaviour when measured by the number of iterations required to reach the predefined thresholds. However, because ESC requires substantially less computation per iteration, it achieved significantly lower overall computational cost.


Tissue motion correction in ULM is particularly suitable for ESC-based image registration. Although ESC may be less effective than some gradient-based methods in escaping local extrema, tissue motion between consecutive ultrasound frames is typically much smaller than the misalignment encountered in multimodal image registration. Consequently, the optimization is generally initialized sufficiently close to the desired solution, reducing the practical impact of local extrema. This characteristic makes ULM motion correction a favourable application scenario for ESC.

Another consideration is the orthogonality requirement of the dither signals. The performance of ESC may degrade when the frequencies of the dither signals are insufficiently separated. For rigid and affine registration, this requirement is readily satisfied because only a small number of parameters are optimized. For B-spline-based registration, although the total number of parameters can be large, only the $4\times4$ control points influencing a local image region require distinct dither frequencies. Therefore, only 32 frequencies are required in the two-dimensional implementation considered in this study, and these frequencies can be reused throughout the control-point grid. This strategy ensures orthogonality among parameters that simultaneously influence the same pixels while maintaining computational efficiency. The weaker orthogonality between neighboring control points may also reduce differences in adjacent control-point displacements and potentially restrict deformation irregularities in the B-spline grid, although this effect requires further investigation.

The short-axis view of the porcine heart was selected for demonstration because the out-of-plane motion was sufficiently small to allow reliable image registration despite the inherent limitations of two-dimensional ultrasound imaging. The maximum in-plane displacement was approximately 1.5 mm, corresponding to approximately 2.3 wavelengths \cite{yan2024transthoracic}. The results shown in Figs. \ref{fig:ULM density map} and \ref{fig:FRC_curves} demonstrate that tissue motion can substantially degrade ULM image quality and resolution. The proposed two-stage ESC-based registration framework effectively corrected tissue motion, resulting in improved vessel delineation and a marked enhancement in FRC-measured resolution. These findings demonstrate the feasibility of ESC-based image registration for ULM motion correction and highlight its potential for computationally efficient super-resolution ultrasound imaging.

The comparison with GD is motivated by the fact that the averaged dynamics of classical ESC approximate gradient descent. However, ESC is not limited to GD-like behavior\cite{Nesic2010UnifyingES}; alternative ESC architectures can be developed to emulate accelerated gradient, momentum-based, Newton, quasi-Newton, and other higher-order optimization methods.

Future work will investigate advanced ESC algorithms and theoretical developments aimed at improving convergence speed and stability. While the present study focused on demonstrating the computational advantages of the basic ESC formulation, advanced variants such as fixed-time ESC \cite{poveda2021nonsmooth,liu2023fast} may provide additional benefits for real-time robotic motion tracking \cite{yan2025online} or ULM imaging \cite{praesius2026realtimeULM}. Future studies will also explore the application of ESC to alternative transformation models, deformation models, and image similarity metrics, as well as its performance in more challenging image registration scenarios involving larger motion magnitudes and more complex tissue deformations.

\section{Conclusion}

This study presented a two-stage image registration framework based on extremum seeking control (ESC) for tissue motion correction in ultrasound localization microscopy (ULM), consisting of affine and B-spline-based non-rigid registration. Simulation and experimental results demonstrated that ESC achieves motion correction performance comparable to gradient descent (GD) while requiring substantially lower computational cost. The proposed framework effectively corrected tissue motion and improved the quality and resolution of reconstructed ULM images. These findings establish ESC as a computationally efficient alternative to conventional gradient-based optimization for image registration and demonstrate its potential for motion correction in ULM. More broadly, the results suggest that ESC may provide an attractive optimization framework for computationally efficient image registration in medical imaging applications involving high-dimensional transformation models. Future work will investigate advanced ESC variants and their application to more complex image registration problems.

\section{Acknowledgement}
We thank the Royal Veterinary College and Prof. Justin Perkins for assistance with the porcine heart extraction. We thank Konstantinos Ntagiantas, Rasheda A Chowdhury, Dimitrios Panagopoulos, Danya Agha-Jaffar, Shengzhe Li and Clara Rodrigo Gonzalez for the help with the \textit{ex vivo} heart experiment.

\bibliographystyle{IEEEtran}
\bibliography{sample,References}

@inproceedings{Nesic2010UnifyingES,
  author    = {Dragan Ne{\v{s}}i{\'c} and Ying Tan and William H. Moase and Chris Manzie},
  title     = {A Unifying Approach to Extremum Seeking: Adaptive Schemes Based on Estimation of Derivatives},
  booktitle = {Proceedings of the 49th IEEE Conference on Decision and Control (CDC)},
  year       = {2010},
  pages      = {4625--4630},
  address    = {Atlanta, GA, USA},
  doi        = {10.1109/CDC.2010.5717929}
}

@article{Scheinker2024Survey,
  author  = {Alexander Scheinker},
  title   = {100 years of extremum seeking: A survey},
  journal = {Automatica},
  volume   = {161},
  pages    = {111481},
  year     = {2024},
  doi      = {10.1016/j.automatica.2023.111481}
}

@book{ariyur-2003,
  title={Real-time optimization by extremum-seeking control},
  author={Ariyur, Kartik B and Krstic, Miroslav},
  year={2003},
  publisher={John Wiley \& Sons}
}

@book{khalil-2002,
  title={Nonlinear systems},
  author={Khalil, Hassan K},
  publisher={Third edition Prentice
Hall, Upper Saddle River, New Jersey},
  year={2002}
}

@INPROCEEDINGS{Teel2001ESC,  author={Teel, A. R. and Popovic, D.},  booktitle={Proceedings of the 2001 American Control Conference},   title={Solving smooth and nonsmooth multivariable extremum seeking problems by the methods of nonlinear programming},   year={2001},  volume={3},   pages={2394-2399}}

@article{rueckert2002nonrigid,
  title={Nonrigid registration using free-form deformations: application to breast MR images},
  author={Rueckert, Daniel and Sonoda, Luke I and Hayes, Carmel and Hill, Derek LG and Leach, Martin O and Hawkes, David J},
  journal={IEEE transactions on medical imaging},
  volume={18},
  number={8},
  pages={712--721},
  year={2002},
  publisher={IEEE}
}

@article{perrot2021so,
  title={So you think you can DAS? A viewpoint on delay-and-sum beamforming},
  author={Perrot, Vincent and Polichetti, Maxime and Varray, Fran{\c{c}}ois and Garcia, Damien},
  journal={Ultrasonics},
  volume={111},
  pages={106309},
  year={2021},
  publisher={Elsevier}
}

@article{christensen2014vivo,
  title={In vivo acoustic super-resolution and super-resolved velocity mapping using microbubbles},
  author={Christensen-Jeffries, Kirsten and Browning, Richard J and Tang, Meng-Xing and Dunsby, Christopher and Eckersley, Robert J},
  journal={IEEE Transactions on Medical Imaging},
  volume={34},
  number={2},
  pages={433--440},
  year={2014},
  publisher={IEEE}
}

@article{errico2015ultrafast,
  title={Ultrafast ultrasound localization microscopy for deep super-resolution vascular imaging},
  author={Errico, Claudia and Pierre, Juliette and Pezet, Sophie and Desailly, Yann and Lenkei, Zsolt and Couture, Olivier and Tanter, Mickael},
  journal={Nature},
  volume={527},
  number={7579},
  pages={499--502},
  year={2015},
  publisher={Nature Publishing Group}
}

@article{praesius2026realtimeULM,
  title={Real-Time Super-Resolution Ultrasound Imaging using the Erythrocytes},
  author={Pr{\ae}sius, Sebastian Kazmarek and J{\o}rgensen, Lasse Thurmann and Jensen, J{\o}rgen Arendt},
  journal={IEEE Transactions on Ultrasonics},
  year={2026},
  publisher={IEEE}
}

@article{harput2018two,
  title={Two-stage motion correction for super-resolution ultrasound imaging in human lower limb},
  author={Harput, Sevan and Christensen-Jeffries, Kirsten and Brown, Jemma and Li, Yuanwei and Williams, Katherine J and Davies, Alun H and Eckersley, Robert J and Dunsby, Christopher and Tang, Meng-Xing},
  journal={IEEE Transactions on Ultrasonics, Ferroelectrics, and Frequency control},
  volume={65},
  number={5},
  pages={803--814},
  year={2018},
  publisher={IEEE}
}

@article{demeulenaere2022coronary,
  title={Coronary flow assessment using 3-dimensional ultrafast ultrasound localization microscopy},
  author={Demeulenaere, Oscar and Sandoval, Zulma and Mateo, Philippe and Dizeux, Alexandre and Villemain, Olivier and Gallet, Romain and Ghaleh, Bijan and Deffieux, Thomas and Dem{\'e}n{\'e}, Charlie and Tanter, Mickael and others},
  journal={Cardiovascular Imaging},
  volume={15},
  number={7},
  pages={1193--1208},
  year={2022},
  publisher={American College of Cardiology Foundation Washington DC}
}

@article{song2023super,
  title={Super-resolution ultrasound microvascular imaging: Is it ready for clinical use?},
  author={Song, Pengfei and Rubin, Jonathan M and Lowerison, Matthew R},
  journal={Zeitschrift f{\"u}r Medizinische Physik},
  volume={33},
  number={3},
  pages={309--323},
  year={2023},
  publisher={Elsevier}
}

@article{sotiras2013deformable,
  title={Deformable medical image registration: A survey},
  author={Sotiras, Aristeidis and Davatzikos, Christos and Paragios, Nikos},
  journal={IEEE transactions on medical imaging},
  volume={32},
  number={7},
  pages={1153--1190},
  year={2013},
  publisher={IEEE}
}

@article{taghavi2021vivo,
  title={In vivo motion correction in super-resolution imaging of rat kidneys},
  author={Taghavi, Iman and Andersen, Sofie Bech and Hoyos, Carlos Armando Villag{\'o}mez and Nielsen, Michael Bachmann and S{\o}rensen, Charlotte Mehlin and Jensen, J{\o}rgen Arendt},
  journal={IEEE Transactions on Ultrasonics, Ferroelectrics, and Frequency Control},
  volume={68},
  number={10},
  pages={3082--3093},
  year={2021},
  publisher={IEEE}
}

@article{hingot2017subwavelength,
  title={Subwavelength motion-correction for ultrafast ultrasound localization microscopy},
  author={Hingot, Vincent and Errico, Claudia and Tanter, Mickael and Couture, Olivier},
  journal={Ultrasonics},
  volume={77},
  pages={17--21},
  year={2017},
  publisher={Elsevier}
}

@article{hingot2021measuring,
  title={Measuring image resolution in ultrasound localization microscopy},
  author={Hingot, Vincent and Chavignon, Arthur and Heiles, Baptiste and Couture, Olivier},
  journal={IEEE transactions on medical imaging},
  volume={40},
  number={12},
  pages={3812--3819},
  year={2021},
  publisher={IEEE}
}

@article{smith2026quantitative,
  title={Quantitative image markers of super-resolution ultrasound},
  author={Smith, Cameron AB and Wilson, Harvey and Yan, Jipeng and Tang, Meng-Xing},
  journal={EBioMedicine},
  volume={124},
  year={2026},
  publisher={Elsevier}
}

@article{demene2021transcranial,
  title={Transcranial ultrafast ultrasound localization microscopy of brain vasculature in patients},
  author={Demen{\'e}, Charlie and Robin, Justine and Dizeux, Alexandre and Heiles, Baptiste and Pernot, Mathieu and Tanter, Mickael and Perren, Fabienne},
  journal={Nature Biomedical Engineering},
  volume={5},
  number={3},
  pages={219--228},
  year={2021},
  publisher={Nature Publishing Group}
}

@article{dencks2025review_super,
  title={Super-resolution ultrasound: from data acquisition and motion correction to localization, tracking, and evaluation},
  author={Dencks, Stefanie and Lowerison, Matthew and Hansen-Shearer, Joseph and Shin, YiRang and Schmitz, Georg and Song, Pengfei and Tang, Meng-Xing},
  journal={IEEE Transactions on Ultrasonics, Ferroelectrics, and Frequency Control},
  year={2025},
  publisher={IEEE}
}

@article{yan2024transthoracic,
  title={Transthoracic ultrasound localization microscopy of myocardial vasculature in patients},
  author={Yan, Jipeng and Huang, Biao and Tonko, Johanna and Toulemonde, Matthieu and Hansen-Shearer, Joseph and Tan, Qingyuan and Riemer, Kai and Ntagiantas, Konstantinos and Chowdhury, Rasheda A and Lambiase, Pier D and others},
  journal={Nature Biomedical Engineering},
  volume={8},
  pages={689--700},
  year={2024},
  publisher={Nature Publishing Group UK London}
}

@article{KRSTIC2000595,
title = {Stability of extremum seeking feedback for general nonlinear dynamic systems},
journal = {Automatica},
volume = {36},
number = {4},
pages = {595-601},
year = {2000},
author = {Miroslav Krstić and Hsin-Hsiung Wang},
}

@inproceedings{wei2024pyramidal,
  title={A Pyramidal Optical Flow Method Based on Dense Sift Feature Maps for Motion Correction in Ultrafast Power Doppler Imaging},
  author={Wei, Xingyue and Huang, Lijie and Lan, Hengrong and Luo, Jianwen},
  booktitle={2024 IEEE International Symposium on Biomedical Imaging (ISBI)},
  pages={1--4},
  year={2024},
  organization={IEEE}
}

@article{TAN2006889,
title = "On non-local stability properties of extremum seeking control",
journal = "Automatica",
volume = "42",
number = "6",
pages = "889 - 903",
year = "2006",
author = "Ying Tan and Dragan Nešić and Iven Mareels"
}

@article{yan2022super,
  title={Super-resolution ultrasound through sparsity-based deconvolution and multi-feature tracking},
  author={Yan, Jipeng and Zhang, Tao and Broughton-Venner, Jacob and Huang, Pintong and Tang, Meng-Xing},
  journal={IEEE Transactions on Medical Imaging},
  volume={41},
  number={8},
  pages={1938--1947},
  year={2022},
  publisher={IEEE}
}

@article{cormier2021dynamic,
  title={Dynamic myocardial ultrasound localization angiography},
  author={Cormier, Philippe and Por{\'e}e, Jonathan and Bourquin, Chlo{\'e} and Provost, Jean},
  journal={IEEE Transactions on Medical Imaging},
  volume={40},
  number={12},
  pages={3379--3388},
  year={2021},
  publisher={IEEE}
}

@article{yan2023fast,
  title={Fast 3D super-resolution ultrasound with adaptive weight-based beamforming},
  author={Yan, Jipeng and Wang, Bingxue and Riemer, Kai and Hansen-Shearer, Joseph and Lerendegui, Marcelo and Toulemonde, Matthieu and Rowlands, Christopher J and Weinberg, Peter D and Tang, Meng-Xing},
  journal={IEEE Transactions on Biomedical Engineering},
  volume={70},
  number={9},
  pages={2752--2761},
  year={2023},
  publisher={IEEE}
}

@article{klein2007evaluation,
  title={Evaluation of optimization methods for nonrigid medical image registration using mutual information and B-splines},
  author={Klein, Stefan and Staring, Marius and Pluim, Josien PW},
  journal={IEEE transactions on image processing},
  volume={16},
  number={12},
  pages={2879--2890},
  year={2007},
  publisher={IEEE}
}

@article{christensen2020super,
  title={Super-resolution ultrasound imaging},
  author={Christensen-Jeffries, Kirsten and Couture, Olivier and Dayton, Paul A and Eldar, Yonina C and Hynynen, Kullervo and Kiessling, Fabian and O'Reilly, Meaghan and Pinton, Gianmarco F and Schmitz, Georg and Tang, Meng-Xing and others},
  journal={Ultrasound in medicine \& biology},
  volume={46},
  number={4},
  pages={865--891},
  year={2020},
  publisher={Elsevier}
}

@inproceedings{liu2023fast,
  title={Fast Extremum Seeking Control for a Class of Generalized Hammerstein Systems with the Knowledge of Relative Degree.},
  author={Liu, Hengchang and Tan, Ying and Bacek, Tomislav and Kulic, Dana and Oetomo, Denny and Manzie, Chris},
  booktitle={ACC},
  pages={2405--2410},
  year={2023}
}

@article{poveda2021nonsmooth,
  title={Nonsmooth extremum seeking control with user-prescribed fixed-time convergence},
  author={Poveda, Jorge I and Krsti{\'c}, Miroslav},
  journal={IEEE Transactions on Automatic Control},
  volume={66},
  number={12},
  pages={6156--6163},
  year={2021},
  publisher={IEEE}
}

@article{yan2025online,
  title={Online 4D ultrasound-Guided robotic tracking enables 3D ultrasound localisation microscopy with large tissue displacements},
  author={Yan, Jipeng and Tan, Qingyuan and Kawara, Shusei and Zhu, Jingwen and Wang, Bingxue and Toulemonde, Matthieu and Liu, Honghai and Tan, Ying and Tang, Meng-Xing},
  journal={IEEE transactions on medical imaging},
  year={2025},
  publisher={IEEE}
}

@article{chen2025survey,
  title={A survey on deep learning in medical image registration: New technologies, uncertainty, evaluation metrics, and beyond},
  author={Chen, Junyu and Liu, Yihao and Wei, Shuwen and Bian, Zhangxing and Subramanian, Shalini and Carass, Aaron and Prince, Jerry L and Du, Yong},
  journal={Medical Image Analysis},
  volume={100},
  pages={103385},
  year={2025},
  publisher={Elsevier}
}

@article{jenkinson2002improved,
  title={Improved optimization for the robust and accurate linear registration and motion correction of brain images},
  author={Jenkinson, Mark and Bannister, Peter and Brady, Michael and Smith, Stephen},
  journal={Neuroimage},
  volume={17},
  number={2},
  pages={825--841},
  year={2002},
  publisher={Elsevier}
}

@article{yoo2002ITK,
  title={Engineering and algorithm design for an image processing API: a technical report on ITK-the insight toolkit},
  author={Yoo, Terry and Chalana, Vikram and Schroeder, Will},
  journal={Studies in health technology and informatics},
  year={2002}
}

\end{document}